\documentclass[aps,prl,twocolumn,a4paper,10pt]{revtex4-1}

\usepackage[T1]{fontenc}    
\usepackage[english]{babel}   
\usepackage[utf8]{inputenc} 
\usepackage{times}

\usepackage{amsmath}      
\usepackage{amssymb}      
\usepackage{bbm}        
\usepackage{mathtools}
\usepackage{simpler-wick}
\usetikzlibrary{tikzmark}
\DeclarePairedDelimiterX\Dirbraket[3]{\langle}{\rangle}%
{#1\,\delimsize\vert\,\mathopen{}#2\,\delimsize\vert\,\mathopen{}#3}

\usepackage{hyperref}
\hypersetup{colorlinks=true, linkcolor=blue, citecolor=blue, urlcolor=blue}

\usepackage{graphicx}     
\DeclareGraphicsExtensions{.pdf}
\usepackage{color}

\usepackage{nicefrac}

\newcommand{\bea}{\begin{eqnarray}}
\newcommand{\eea}{\end{eqnarray}}

\newcommand{\mbf}{\mathbf}

\definecolor{mag}{rgb}{1., 0., 1.}

\definecolor{darkGreen}{rgb}{0.0, 0.5, 0.0}

\begin{document}

\title{Evidence of attraction between charge-carriers in a doped Mott insulator}

\author{Emil Blomquist$^{1}$ and Johan~Carlstr\"om$^{2}$ }
\affiliation{$^{1}$Department of Physics, Royal Institute of Technology, Stockholm, SE-106 91, Sweden}
\affiliation{$^{2}$Department of Physics, Stockholm University, 106 91 Stockholm, Sweden}
\date{\today}

\begin{abstract}

Recent progress in optically trapped ultracold atomic gases is now making it possible to access microscopic observables in doped Mott insulators, which are the parent states of high-temperature superconductors. This makes it possible to address longstanding questions about the temperature scales at which attraction between charge carriers are present, and their mechanism. Controllable theoretical results for this problem are not available at low temperature due to the sign problem. In this work, we overcome this longstanding obstacle by employing worm-algorithm Monte Carlo, which allows us to obtain completely unbiased results for two charge carriers in a Mott insulator. Our method gives access to lower temperatures than what is currently possible in experiments, and provides evidence for attraction between dopants at a temperature scale that is now feasible in ultracold atomic systems. We also report on spin-correlations in the presence of charge carriers, which are directly comparable to experiments. 
\end{abstract}
\maketitle

Ultracold atomic gases have provided an alternative path to exploring the physics of high-temperature superconductors by emulation of a doped Mott insulator.  
This system, with charge carriers propagating on an antiferromagnetic background, is widely regarded as the parent state of high-temperature superconductivity \cite{RevModPhys.78.17}. In this setting, the dopants form quasi-particles in the form of polarons, as a result of competition between kinetic energy and super-exchange processes. The carrier can reduce its kinetic energy by delocalizing, but this naturally distorts the spin background. Minimizing the super-exchange energy leads to a state with strong antiferromagnetic correlations, which increases the kinetic energy. The polaron emerges as the best compromise.

It is generally believed that pairing in the high-temperature superconductors occurs as a result of an attractive interaction between charge carriers that is mediated by the spin-background. The precise nature of the microscopic mechanism remains debated, however. A circumstance which complicates this question is that superconductivity does not emerge from a conventional metallic state, but rather, from a non-Fermi liquid \cite{RevModPhys.78.17}.

The search for a pairing mechanism has lead to considerable interest in how doping alters spin-correlations in Mott insulators. For example, the resonating valence bond (RVB) theory claims that the spins form a superposition of singlets, providing a better compromise between kinetic and magnetic energy \cite{ANDERSON1196}. If such a mechanism is at play, then it should have implications for the spin-background, possibly even above the onset of superconductivity.

Another point of contention is whether fermions form pairs before the onset of superconductivity, such that these merely condense at the critical temperature \cite{Emery1995,PhysRevB.55.3173}.
 The conjecture of preformed pairs seems to be invited by the scale of the super-exchange parameter $J$, which in the high-temperature superconductors corresponds to approximately $1500$K \cite{RevModPhys.70.897,PhysRevB.41.11068}.
This raises questions about the temperature range where attraction between charge carriers can be observed.
In light of these considerations, spin- and charge-correlations have become the focal point of ultracold-atoms experiments.

Despite intense research, both the pairing mechanism and the nature of the normal state remain open questions, chiefly, as a consequence of a lack of theoretical methods that can reliably address strongly correlated fermions. The Hubbard model--which is believed to capture the essential physics of cuprates--has been investigated with a number of approximative many-body techniques, including density-matrix renormalization-group theory (DMRG) \cite{PhysRevLett.69.2863}, dynamic mean-field theory \cite{PhysRevLett.62.324, PhysRevB.75.045118, PhysRevB.77.033101,PhysRevLett.106.047004,PhysRevLett.110.216405,RevModPhys.77.1027} and auxiliary field quantum Monte Carlo \cite{PhysRevB.39.839,PhysRevLett.62.591,PhysRevB.40.506,afqmc}. Comparing these methods reveals discrepancies, which lead to the conclusion that the phase diagrams of strongly correlated systems are highly sensitive to uncontrolled approximations \cite{2006cond.mat.10710S,PhysRevX.5.041041}. To date, there are few unbiased numerical techniques for strongly correlated fermions. Numerical linked cluster expansion has provided the equation of state as well as spin-correlations to exceptional precision \cite{PhysRevLett.97.187202,PhysRevE.89.063301}, but has thus far not given any insights into polaron physics or superconductivity. 
Diagrammatic Monte Carlo techniques \cite{VANHOUCKE201095} can resolve pairing in the fermi-liquid regime \cite{0295-5075-110-5-57001}. 
Recently, this method has also been extended to strongly correlated systems, though currently published results are limited to relatively high temperatures \cite{0953-8984-29-38-385602, PhysRevB.97.075119, carlstrm2019strongcoupling}.
However, for small systems, exact diagonalization and the Lanczos technique do provide indications of attraction between charge carriers \cite{PhysRevB.49.12318}.

The lack of theoretical methods for correlated fermions has motivated an intense effort to develop experimental techniques with access to microscopic observables.
With the advent of quantum gas microscopy, imaging of entangled many-body states is now possible at the level of single-site resolution \cite{Gross995}. Using ultracold atomic gases, Mott-insulating states can be created and cooled to the point where antiferromagnetic correlations become significant \cite{Mazurenko2017}.

Recent experimental work has focused on spin-correlations in doped systems \cite{Cheuk1260,Parsons1253}. In particular, this has resulted in the first observation of the internal structure of a polaron on an antiferromagnetic background, revealing microscopic details about the cloud of spins surrounding the carrier \cite{Koepsell2019}. The search for correlations between dopants and thus signs of effective interactions has also been initiated \cite{Chiu251}.

In this work, we use worm-algorithm Monte Carlo \cite{Prokofev1998} (WAMC) to extract completely unbiased charge and spin correlations in a doped Mott insulator, in the presence of thermal fluctuations. Our findings indicate that attraction between carriers is present at an energy scale corresponding to approximately $700$K, significantly above the critical temperature. This is a regime that could realistically be achieved in ultra-cold atomic gases in the near future. We also report on spin-correlations in the presence of two interacting dopants, finding that the delocalization of multiple carriers can explain the reversal of correlators seen in recent experiments \cite{Chiu251,Parsons1253}.

WAMC is a technique that allows extremely efficient sampling of world-lines, from which we obtain diagonal elements of the density matrix at thermal equilibrium. Our observables are therefore not subject to any bias, since they are exactly represented by the distribution of world lines. This sets our method apart from previous works, which are either based on uncontrolled approximations, or are confined to small systems, where finite size effects are considerable.  
WAMC is mainly applied to bosons due to the fermionic sign. However, for Gutzwiller-projected theories, the sign problem is only extensive in the number of charge carriers as opposed to the system size. Using a highly efficient sampling protocol, it is possible to address a small number of dopants in a system that is sufficiently large to avoid finite-size effects. Previously, this method has been used to extract spectral properties of a single carrier \cite{PhysRevB.64.033101}. More recently, it has also provided the real-space structure of a single polaron \cite{blomquist2019ab}.

In this work, we represent charge carriers and spins as world-lines in space and imaginary time. We use a $20\times 20$ lattice with periodic boundary conditions to prevent finite-size effects. We use separate worms for the spin and charge sectors \cite{Prokofev1998}. The former can wind in imaginary time, which alters the total spin in the system so that this sector is in the grand canonical ensemble.
The worm corresponding to charge cannot wind, keeping the number of carriers to two at all times (in the partition function sector).
By generating contributions to the trace of the density matrix at thermal equilibrium, we obtain access to essentially arbitrary operator expectation values, including spin and charge correlators.
To confirm the accuracy of this method, we provide benchmarks against exact diagonalization, see supplementary material at [URL will be inserted by publisher].

We describe the system using the $ t $-$ J $ model \cite{spalek}
\begin{equation}
  \label{eq:H_tJ}
  \begin{split}
    \hat H
    &=
    - t \sum _{\langle ij \rangle, \,\sigma}
    \hat c_{i \sigma}^\dagger (1 - \hat n_{i \bar \sigma})
    \hat c_{j \sigma} (1 - \hat n_{j \bar \sigma}) \\
    &+ J \sum _{\langle ij \rangle}
    \left(
    \hat{\mbf S}_i \cdot \hat{\mbf S}_j
    -
    \frac{\hat{n}_i\hat{n}_j}{4}
    \right)  \,,\; \bar{\sigma}=-\sigma,
  \end{split}
\end{equation}
which effectively captures the low energy physics of the Hubbard model \cite{Hubbard238} in the case of large onsite repulsion. Here $ \hat c_{i \sigma} $, $ \hat c^\dagger_{i \sigma} $ denote annihilation and creation operators of a spin $ \sigma$ electron at the site $ i $. The total particle number on the site $i$ is given by $ \hat n_{i}=\sum_\sigma \hat n_{i,\sigma}$ where $ \hat n_{i,\sigma}= \hat c^\dagger_{i \sigma} \hat c_{i \sigma} $.
Two energy scales thus describe the model (\ref{eq:H_tJ}). The kinetic energy is due to hole propagation, and is proportional to $t$. The super-exchange energy originates in the virtual creation and annihilation of doublon-hole pairs and is proportional to $J=4t^2/U$, where $U$ is the onsite repulsion.

In the low mobility limit, where $t/J\to 0$, the attraction between two carriers can be understood from a simple broken-bond picture: An isolated hole negates $4$ super-exchange interactions. When two carriers share a link, the number of broken bonds is reduced from $8$ to $7$, saving some magnetic energy.
For realistic model parameters, where $t/J$ is not vanishing, this picture becomes overly crude, however. Binding the two holes together impairs delocalization, resulting in competition between kinetic and exchange-mediated attraction. Therefore, it is expected that the binding energy decreases with increasing $t/J $, to possibly vanish at some critical value $(t/J)_c$ \cite{RevModPhys.66.763}.
This picture is corroborated by zero-temperature calculations on small clusters based on the Lanczos algorithm, where it is found that with increasing $t/J$, the binding energy decreases, while the typical separation grows. However, being limited to systems of up to $26$ lattice site, these results cannot resolve bound pairs where the distance between carriers is large compared to the system size \cite{PhysRevB.49.12318}.
 DMRG calculations on notably larger clusters--up to $10\times 7$ sites--also indicate attraction at zero temperature \cite{White_1997}.

At realistic values of $t/J$, cluster calculations give a relatively small amplitude for finding the holes on neighboring sites, stressing the inadequacy of the broken-bond model \cite{PhysRevB.49.12318}. Meanwhile, the delocalization of a charge carrier gives rise to frustration due to competition between kinetic and magnetic processes. This has motivated the suggestion that frustrated bonds may mediate attraction between carriers \cite{White_1997}.

\begin{figure}[!htb]
  \includegraphics[width=\linewidth]{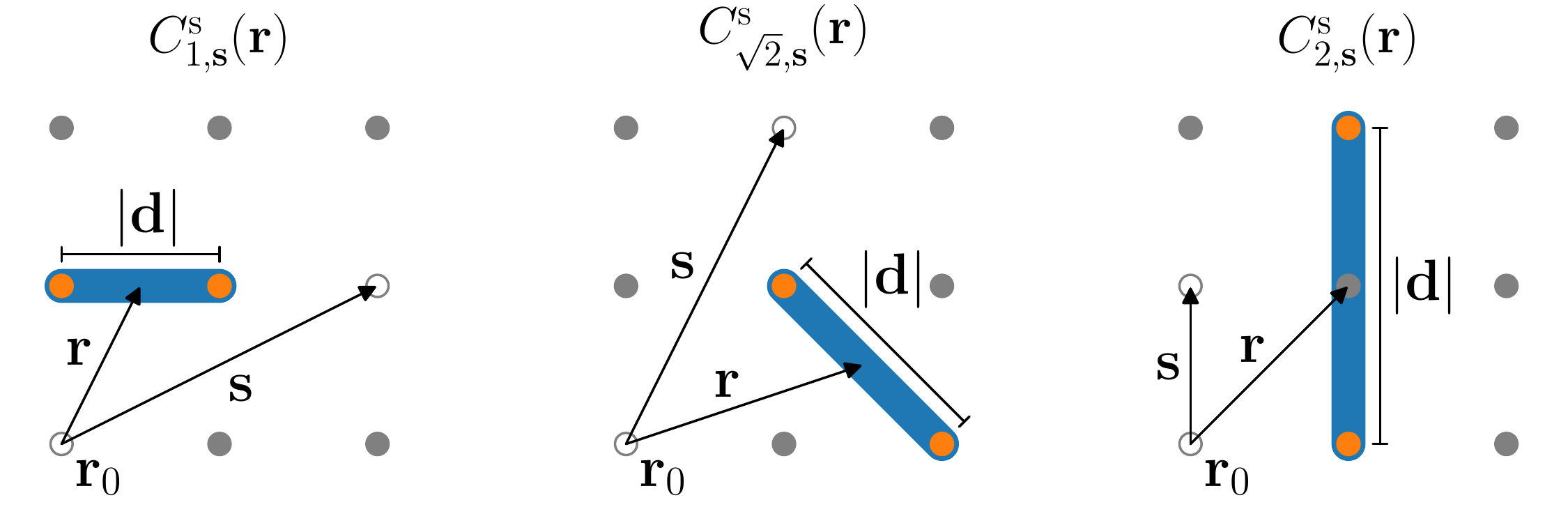}
  \caption{
    \textbf{Illustration of spin-spin correlators}. The spin-spin correlators $ C^\mathrm{s}_{\mathbf{|d|}, \mathbf{s}}(\mathbf{r}) $ defined in Eq.\ (\ref{eq:spin-corr}) illustrated in the case of $ | \mathbf d | = 1 $, $ | \mathbf d | = \sqrt 2 $, and $ | \mathbf d | = 2 $. Here, the bond distance $ \mathbf r $ is defined as the distance from one of the carriers to the point between the spins for which the correlation is considered and $ \mathbf s $ is the distance to the second carrier.
  }
  \label{fig:illustration}
\end{figure}


To obtain indications of attraction between charge carriers that are directly comparable to experiments, we calculate the hole-hole correlator---in a frame of reference where one hole is always located at the origin---which takes the form
\begin{equation}
  C^\mathrm{h}(\mathbf{s})
  =
  (N-1)
  \big \langle
  \hat n^\mathrm{h}_{\mathbf{0}}
  \hat n^\mathrm{h}_{\mathbf{s}}
  \big \rangle - 1 \,.
  \label{Ch}
\end{equation}
Here $ \hat n^\mathrm{h}_\mathbf{r} =1-\hat{n}_\mathbf{r}$, is the hole-number operator at site $ \mathbf{r} $, $ \mathbf s $ is the distance between the two holes and $N$ is the number of lattice sites. Since the spin background is expected to mediate attraction between carriers, we also calculate spin-correlations of the form
\begin{equation}
  \label{eq:spin-corr}
  C^\mathrm{s}_{\mathbf{|d|}, \mathbf{s}}(\mathbf{r})
  =
  4
  \frac{
    \sum \limits_{\mathbf{r}_0}
    \big \langle
      \hat n^\mathrm{h}_{\mathbf{r}_0}
      \hat n^\mathrm{h}_{\mathbf{r}_0 + \mathbf{s}}
      \hat S^z_{\mathbf{r}_0 + \mathbf{r} - \mathbf{d}/2}
      \hat S^z_{\mathbf{r}_0 + \mathbf{r} + \mathbf{d}/2}
    \big \rangle
  }{
    \sum \limits_{\mathbf{r}_0}
    \big \langle
      \hat n^\mathrm{h}_{\mathbf{r}_0}
      \hat n^\mathrm{h}_{\mathbf{r}_0 + \mathbf{s}}
    \big \rangle
  } \,.
\end{equation}
Here $ \mathbf{r}_0 $ and $ \mathbf{r}_0 + \mathbf s $ are the positions of the two carriers, and similarly $ \mathbf{r}_0 + \mathbf{r} \pm \mathbf{d}/2 $ are the positions of the two spins, located a distance $ \mathbf d $ from one another. We will consider the cases $ | \mathbf d | = 1 $ (nearest-neighbor), $ | \mathbf d | = \sqrt 2 $ (next-nearest-neighbor), and $ | \mathbf d | = 2 $ (next-next-nearest-neighbor). These correlators are illustrated in Fig.\ \ref{fig:illustration}.

\begin{figure}[!htb]
  \includegraphics[width=\linewidth]{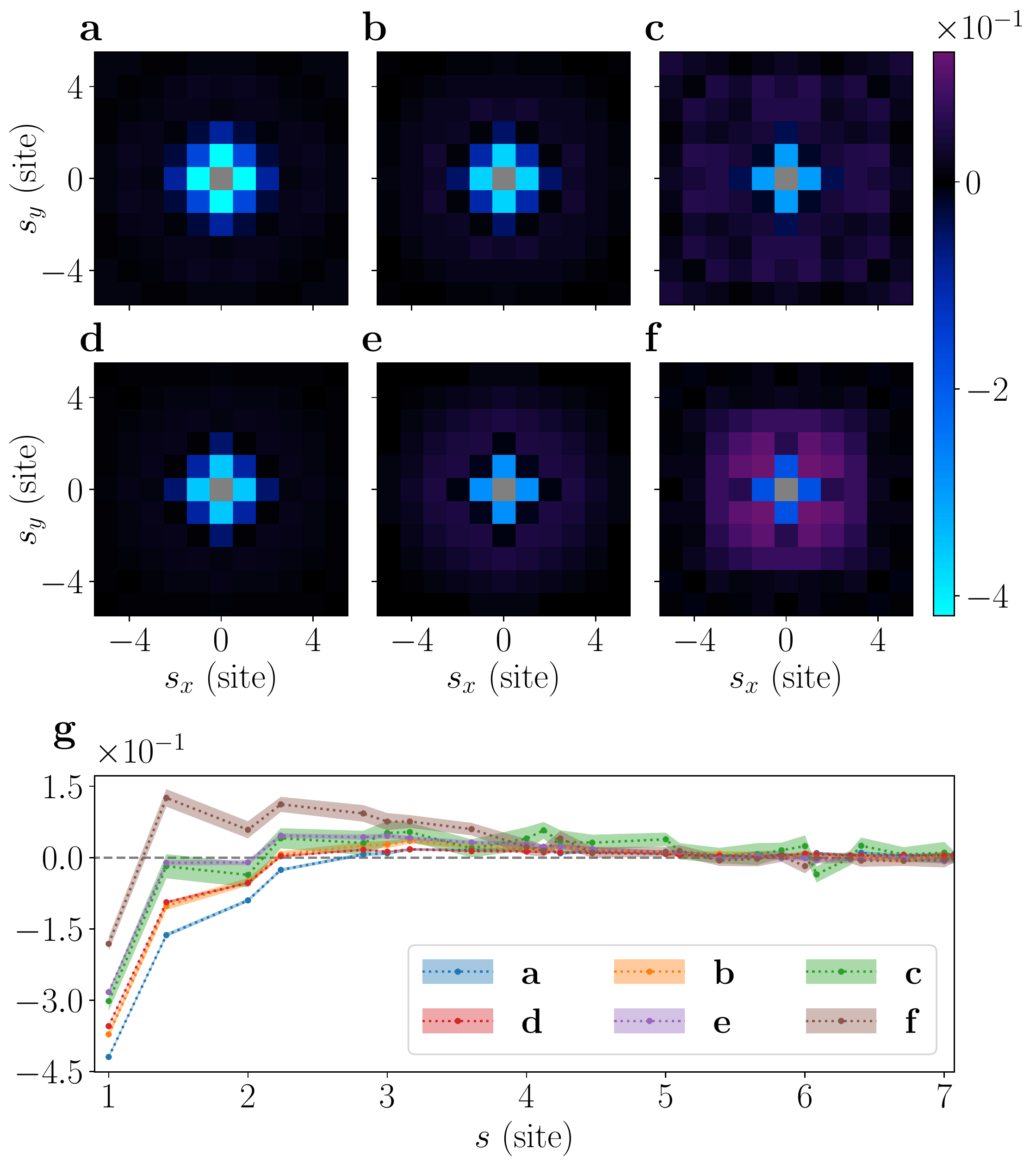}
  \caption{
    \textbf{Hole-hole correlations.}
    (\textbf{a}-\textbf{c}) show contour plots of the correlator $C^\mathrm{h}(\mathbf{s})$ (Eq. \ref{Ch}) for the case $ t/J = 2 $ and $ \beta J = 2.1,\; 2.4, \;2.7$ respectively. The radial components of these are given in (\textbf{g}). For this parameter set, we observe a peak in the probability distribution at small separation which is visible already at temperatures corresponding to $ J \beta = 2.1 $. In (\textbf{a}) the effect is small, with a peak value of $ C^\text{h} \approx (2.3 \pm 0.27) \times 10^{-2} $ at $ s \approx 4.12 $. Decreasing the temperature bolsters the effect, and in (\textbf{c}) we observe a maximum value of $ C^\text{h} \approx (6.0 \pm 1.7) \times 10^{-2} $, also at a separation of $ s \approx 4.12 $. This scenario corresponds to $ U = 8t $ in the Hubbard model, which is the parameter set that was realized in the experiment \cite{Chiu251}.
    \\
    Slightly decreasing $t/J$ increases the height of the peak of $C^\text{h}$.
    In (\textbf{d}-\textbf{f}), contour plots of $C^\mathrm{h}(\mathbf{s})$ are given for $t/J=5/3$ and $\beta J =   2.1,\;2.4, \;2.7 $ respectively. The corresponding radial components are again shown in (\textbf{g}).
    At the lowest temperature (\textbf{f}), the peak in $ C^\text{h} $ is found when the holes are next-nearest neighbors.
  }
  \label{panel}
\end{figure}

In Fig. \ref{panel}, we show examples of hole-hole correlations for the cases $t/J=2$ and $t/J=5/3$. These reveal an attraction between the carriers which is manifested in a peak in the correlator $ C^\text{h} $ (Eq. \ref{Ch}) at small separation.

The scenario when $t/J=2$ corresponds to $U=8t$ in the Hubbard model, i.e. where the onsite repulsion is equal to the band width. For this parameter set we observe a peak in $ C^\text{h} $ which is present at $ J\beta = 2.1 $ (Fig. \ref{panel}, \textbf{a}).
For comparison, the energy scale of the super-exchange in cuprate superconductors has been estimated to $J=128\pm5 $meV, corresponding to approximately $1500$K. This suggests an onset of weak attraction at a temperature equivalent to approximately $700$K, though it should be stressed that the ratio of the kinetic and magnetic energy scales in the cuprates is somewhat larger, with estimates at $t/J\approx 3.3 $ \cite{RevModPhys.70.897,PhysRevB.41.11068}.
This parameter set was recently realized in an optically trapped ultracold atomic gas. Examination of charge-correlations at the temperature $ J\beta \approx 1.53 $ did not reveal any signs of attraction between carriers, consistent with our results \cite{Chiu251}.

Zero-temperature estimates of the binding energy obtained from the Lanczos algorithm and DMRG--with significant uncertainty due to finite size effects--give results in the span $ 0.6 \ge \epsilon_{\text{binding}}/t\ge 0.15 $ when $ t/J=2 $. In a temperature range where $\beta t=4.2-5.4$, this energy scale is sufficient to impose significant correlations, and the fact that these only reach $ C^\text{h} \approx 6\% $ suggests that thermal fluctuations renormalize the interactions between charge carriers.
This can partially be attributed to the loss of the magnetic correlations that mediate attraction. However, it is also the case that spin fluctuations increase the delocalization of the carrier, which also works against pair formation. The latter effect is manifested in the fact that the kinetic energy of a dopant may actually increase as the temperature is lowered \cite{blomquist2019ab}.

The fact that fluctuations renormalize interactions--even to the point where attraction vanishes--suggests that ground state results provide limited guidance regarding the temperature ranges where we can expect to observe attraction in ultracold atomic gases, where strong temperature effects are ubiquitous.

Decreasing the scale of the kinetic energy to $t/J=5/3$ suppresses delocalization. The peak in $ C^\text{h} $ sets in at $J\beta\approx 1.8$ and is now larger at comparable temperatures. Eventually, the correlator attains its maximum when the carriers are next-nearest neighbors (Fig. \ref{panel}, \textbf{f}).

\begin{figure}[!htb]
  \includegraphics[width=\linewidth]{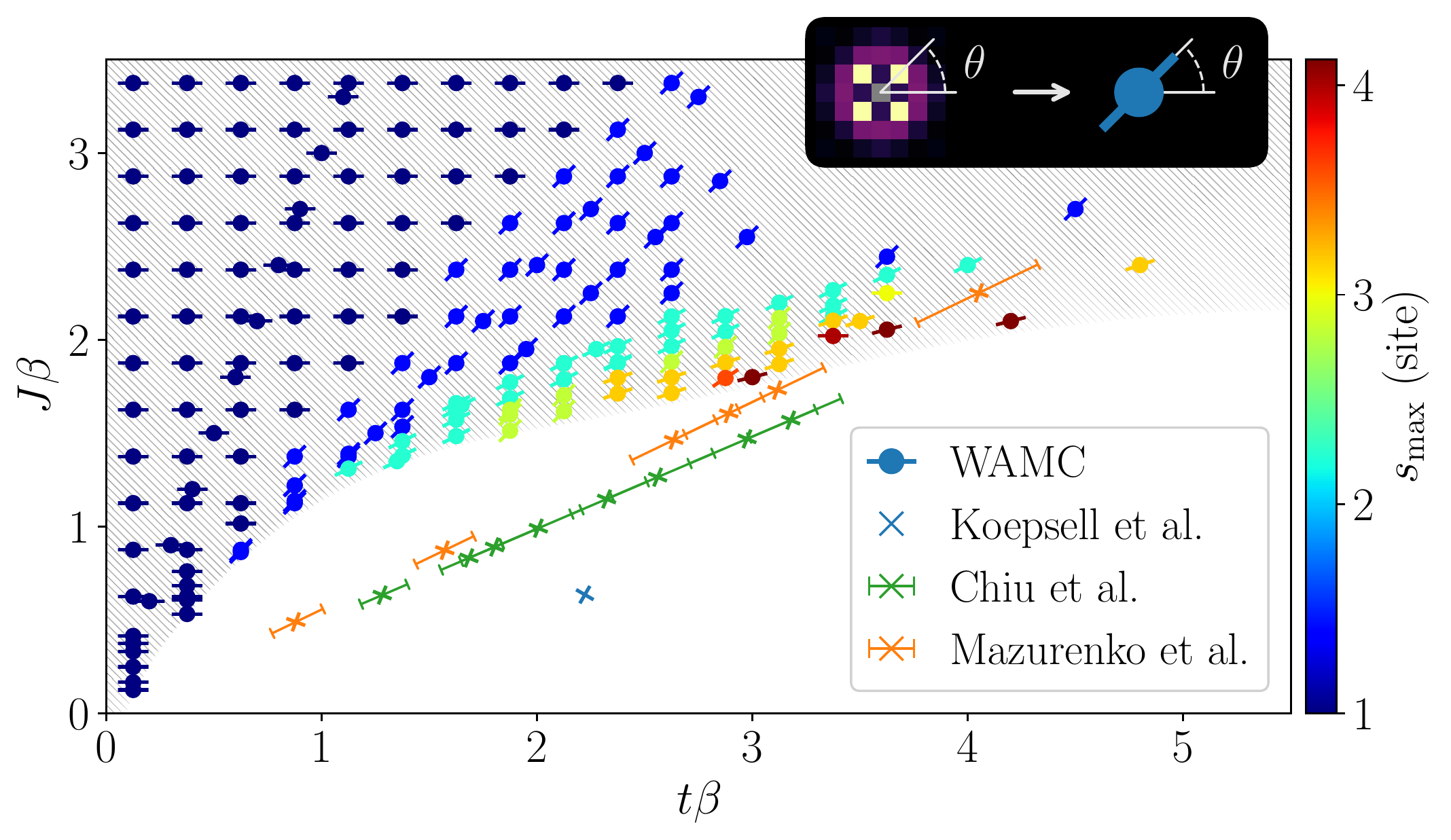}
  \caption{
    \textbf{Parameter regions with attraction.}
    Discs denote parameters for which the correlator $ C^\text{h} $ exhibits a peak at a carrier separation $\mathbf{s}_\text{max}$, which indicates attraction. The bar intersecting the disc gives the orientation of $\mathbf{s}_\text{max}$, while the color gives its magnitude according to the legend.
    The shaded region is a guide to the eye. For comparison, we also indicate the parameter sets of three recent experiments.
    In Koepsell et al.\ \cite{Koepsell2019}, the internal structure of a single polaron was mapped out using quantum gas microscopy.
    In Chiu et al.\ \cite{Chiu251}, correlations between dopants in an ultracold atomic gas were examined.
    In Mazurenko et al.\ \cite{Mazurenko2017}, spin-correlations in the Hubbard model were examined, though this parameter set does strictly speaking not correspond to a Mott insulator.
  }
  \label{fig:phase_diagram}
\end{figure}

By generating hole-hole correlations for a wide range of parameters and systematically testing whether $ C^\text{h} $ possesses a maximum (within two standard deviations, see supplementary information at  [URL will be inserted by publisher]  for details), we obtain the phase diagram shown in Fig. \ref{fig:phase_diagram}. To put this data into context, we have included the model parameters for three recent experiments. The polaron studied in \cite{Koepsell2019} is situated far from the region of attraction, owing to a relatively high temperature and also large onsite repulsion.
The experiment on correlations between dopants reported in \cite{Chiu251} is, however, closer to the onset of attraction.
For this parameter set, our results indicate that a peak in $ C^\text{h} $ appears at $ J\beta \approx 2.1 $.
Even lower temperatures are achievable, as exemplified by \cite{Mazurenko2017}, were long-range spin-correlations were reported at
$T/t\approx 0.25$.
The phase diagram confirms that the attraction is sensitive to the ratio $ t/J $. This is also the case at zero temperature, and stems from competition between delocalization and binding of the carriers \cite{RevModPhys.66.763}.

\begin{figure}[!htb]
  \includegraphics[width=\linewidth]{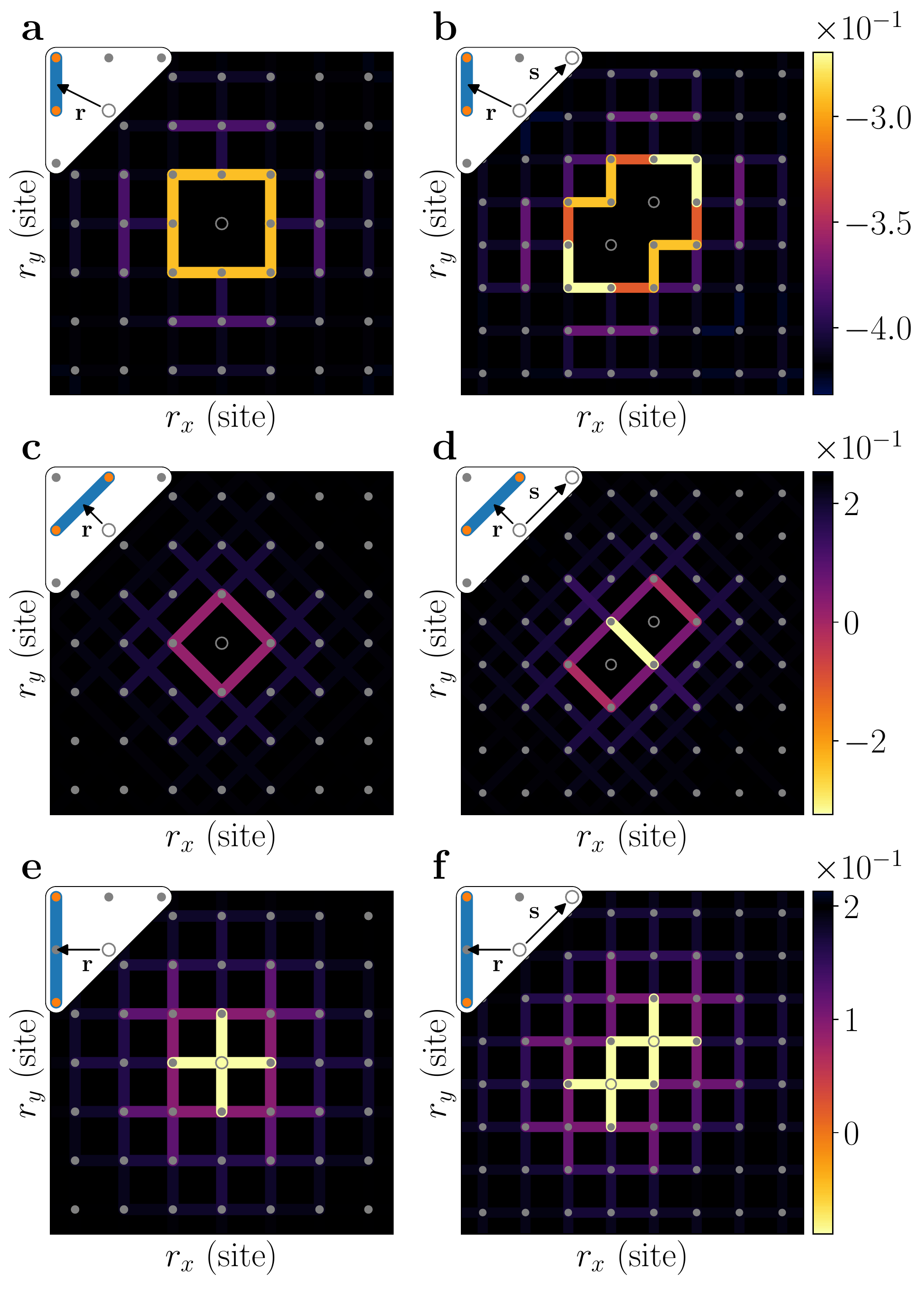}
  \caption{
    \textbf{Spin correlations $ C^\mathrm{s}_{|\mathbf d|, \mathbf s}(\mathbf r) $} in the proximity of carriers. Left column depicts a single hole, while right column shows two dopants which are next nearest neighbors.
    The top row (\textbf{a}, \textbf{b}) gives
    $C^\mathrm{s}_{|\mathbf d|=1, \mathbf s}(\mathbf r) $,
    middle row (\textbf{c}, \textbf{d}) gives $C^\mathrm{s}_{|\mathbf d|=\sqrt{2}, \mathbf s}(\mathbf r) $,
    and bottom row (\textbf{e}, \textbf{f}) gives $C^\mathrm{s}_{|\mathbf d|=2, \mathbf s}(\mathbf r) $.
    Model parameters are $ J \beta = 2.4 $ and $ t/J = 2 $.
    Additional examples of spin-correlations are provided in the supplementary material at  [URL will be inserted by publisher]. Animations can be found here \cite{youtube}.  
  }
  \label{fig:ssc_huge_t}
\end{figure}

 In Fig. \ref{fig:ssc_huge_t}, we show spin-correlators (Eq. \ref{eq:spin-corr}) for the case $ t/J = 2 $ and $ J\beta = 2.4 $, which is in the attractive regime.
 For a single carrier (\textbf{a}, \textbf{c}, \textbf{e}) we observe a cloud of distorted spin-correlations as a result of competition between delocalization and super-exchange, in line with \cite{Koepsell2019, 
 blomquist2019ab,PhysRevB.99.224422}. This includes a suppression of nearest neighbor correlators (\textbf{a}) as well as the reversal of next-next-nearest neighbor correlations across the carrier (\textbf{c}).

\begin{figure}[!htb]
  \includegraphics[width=\linewidth]{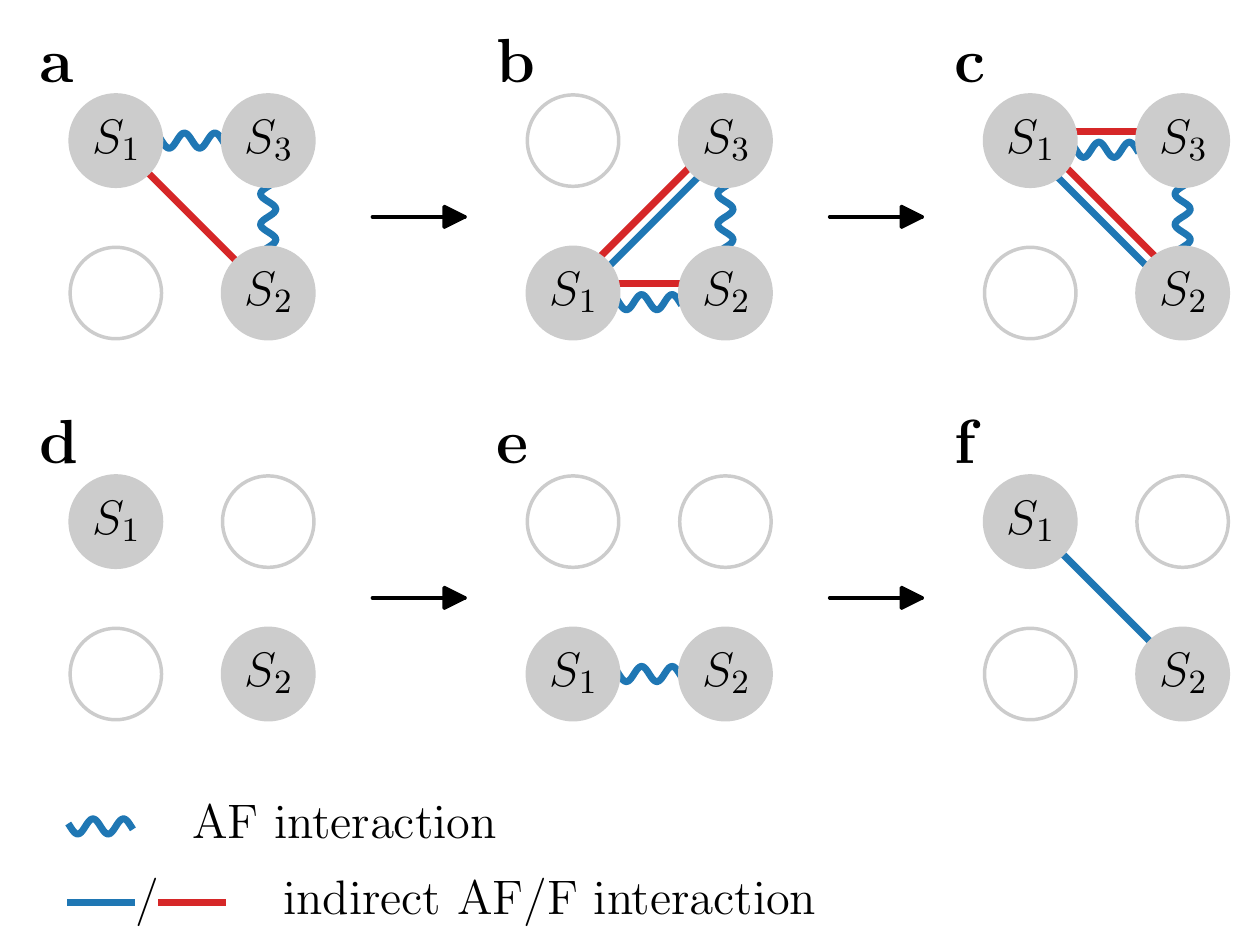}
  \caption{
  \textbf{Super-exchange and kinetic-magnetic interaction}. In the presence of a single hole (\textbf{a}), the spins $S_1$ and $S_2$ interact indirectly via super-exchange that is mediated by $S_3$. This gives rise to an effective ferromagnetic interaction. However, delocalization of the hole (\textbf{a}-\textbf{c}) brings $S_1$ and $S_2$ into direct contact, where interactions are antiferromagnetic. We refer to this as kinetic-magnetic interaction. The result is a high level of frustration and almost vanishing spin-correlations near the carrier.
  When two holes are present (\textbf{d}-\textbf{f}), only kinetic-magnetic interaction is present so that $S_1$ and $S_2$ become anti-correlated.
  }
  \label{fig:hop}
\end{figure}

 In the proximity of the carrier, the next-nearest neighbor correlators (\textbf{b}) are strongly suppressed due to competing processes: When the dopant hops a single lattice spacing, spins that were previously next-nearest neighbors--and therefore would be correlated--are brought into direct contact, where the interaction is antiferromagnetic. This leads to a highly frustrated state with competition between a kinetic-magnetic interaction and super-exchange, see Fig. \ref{fig:hop}, (\textbf{a}-\textbf{c}).

When two carriers are placed as next-nearest neighbors, the frustration is alleviated. The primary interaction between the two spins which are situated next to both dopants is now kinetic-magnetic (Fig. \ref{fig:hop}, \textbf{d}-\textbf{f}), causing remarkably strong anti-correlation as seen in Fig.  \ref{fig:ssc_huge_t} (\textbf{d}).
This result is consistent with magnetic properties, which have been observed in several recent experiments: When doping a Mott insulator, it is found that next-nearest neighbor spin-correlations cross over from positive to negative at a carrier density of approximately $20\%$ \cite{Chiu251,Parsons1253}. Notably, this cross over is not predicted by the structure of a single polaron, and the generation of strong anti-correlations must be understood as a multiple charge-carrier phenomenon. The alleviation of frustration has been identified as a possible source of attraction between carriers in doped Mott insulators \cite{White_1997}.

In conclusion, we present high-precision data for spin- and charge-correlations in a Mott insulator with two dopants at finite temperature, obtained via worm-algorithm Monte Carlo.
To the best of our knowledge, this is the first time that unbiased theoretical results are reported for this problem.
Our findings indicate that the interactions are renormalized by thermal fluctuations, and that attraction sets in at $J\beta \approx 2.1 $ when the onsite repulsion is equal to the bandwidth (i.e. $U=8t$).
This is a temperature range that is now experimentally feasible, and we argue that real progress in understanding this longstanding problem is now possible using optically trapped ultracold atoms and high-precision quantum Monte Carlo simulations in tandem. 

Our technique gives access to lower temperatures than what is possible in experiments, allowing us to provide a phase diagram outlining parameter regions where attraction is present. We make detailed predictions about spin-correlation in the proximity of the two carriers that are directly comparable to quantum gas microscopy and thus serve as a natural bench mark for future experiments.

{\it Acknowledgments---}
This work was supported by the Swedish Research Council (VR) through grants No 2018-03882, 642-2013-7837 and 2018-03659. It was also supported by G{\aa}l\"ostiftelsen, through a travel stipend. The computations were enabled by resources provided by the Swedish National Infrastructure for Computing (SNIC) at the National Supercomputer Centre in Link\"oping  partially funded by the Swedish Research Council through grant agreement no. 2018-05973.
\bibliography{biblio}

\end{document}


\title{Supplementary Information}


\maketitle



\subsection{Worm-algorithm Monte Carlo}

Using a continuous-time worm-algorithm Monte Carlo (WAMC), we simulate a pair of holes in the $t$-$J$ model \cite{spalek} on a quadratic lattice with $ 20 \times 20 $ sites and periodic boundary conditions. The WAMC method effectively samples world-line configurations in real-space and imaginary-time by combining the partition function sector and Green's function sector: In order to go from one partition function world-line configuration to another, one needs to pass through the Green's function sector \cite{Prokofev1998}.
We use two different worms to represent spins and holes. The former is allowed to wind in imaginary-time such that the total spin can vary. Contrary, the hole worm is not, and the number of holes in the partition function sector is therefore kept fixed. All data presented are extracted purely from the sampled partition function configurations. In particular, the kinetic energy presented later in this supplementary information is obtained by counting the kinks involving the hole worm \cite{Sellin1173323}.

\subsection{Sign problem}

When simulating two holes in the $t$-$J$ model using WAMC, there are two types of processes contributing to the fermionic sign problem: exchange of indistinguishable spins and exchange of holes. The former enters first at order $ t^2 J^3 $ \cite{PhysRevB.64.033101} and the latter at order $ t^4 $. When such a process is present in a world-line configuration, the configuration weight $ p $ might turn negative. Hence the name ``sign problem''.

\begin{figure}[!htb]
  \includegraphics[width=\linewidth]{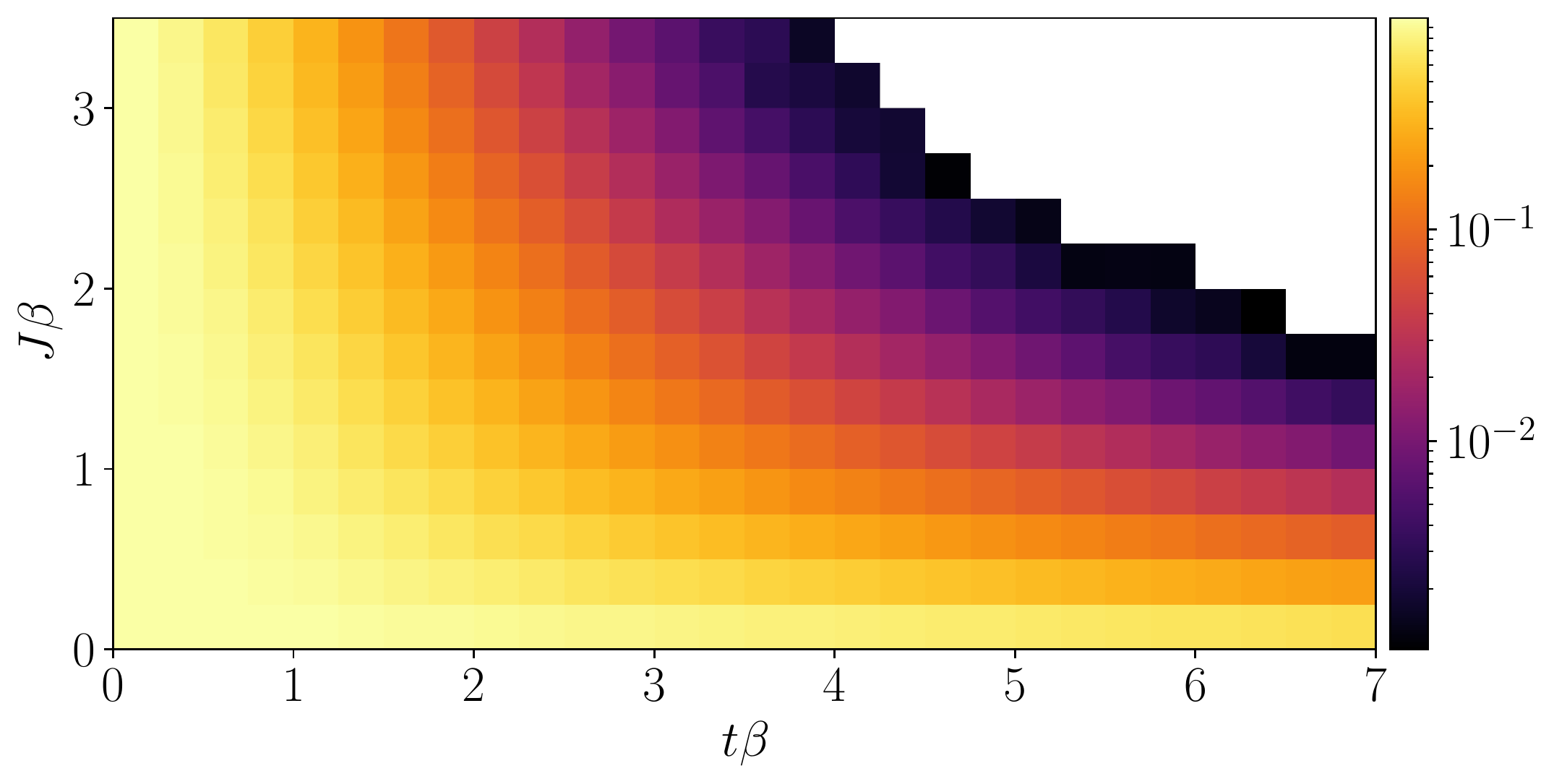}
  \caption{
    \textbf{Average sign.}
    This plot shows the exponential decay of $ \langle s \rangle' $ with increased values of $ t \beta $ and $ J \beta $. Only values $ \langle s \rangle' > 10^{-3} $ are shown due to an otherwise poor SNR.
  }
  \label{fig:sign_map}
\end{figure}

In order to account for negative configuration weights in WAMC, we factor the weight $ p $ into its sign $ s = \pm 1 $ and magnitude $ |p| $, i.e., $ p = s |p| $.
Then, when computing the acceptance ration, $ p $ is exchanged in favor of $ |p| $, and every sampled quantity $ x $ is assigned the sign $ s $ of the world-line configuration. By performing this substitution, the acquired expectation value $ \langle x \rangle' $ is not that of the original fermionic system. However, the latter one is obtained through $ \langle x \rangle = \langle sx \rangle' / \langle s \rangle' $ \cite{PhysRevLett.94.170201}.
%
%
%
This way of circumventing negative configuration weights gives rise to an exponential decrease in signal-to-noise ratio (SNR) and is the hallmark of the sign problem. The decrease of SNR can, in part, be explained by the exponential decay of the denominator $ \langle s \rangle' $ with --- in our case --- increased values of $ t\beta $ and $ J\beta $. This decay of $ \langle s \rangle' $ is illustrated in \Figref{fig:sign_map}.

\subsection{Determining hole-hole attraction}

\begin{figure}[!htb]
  \includegraphics[width=\linewidth]{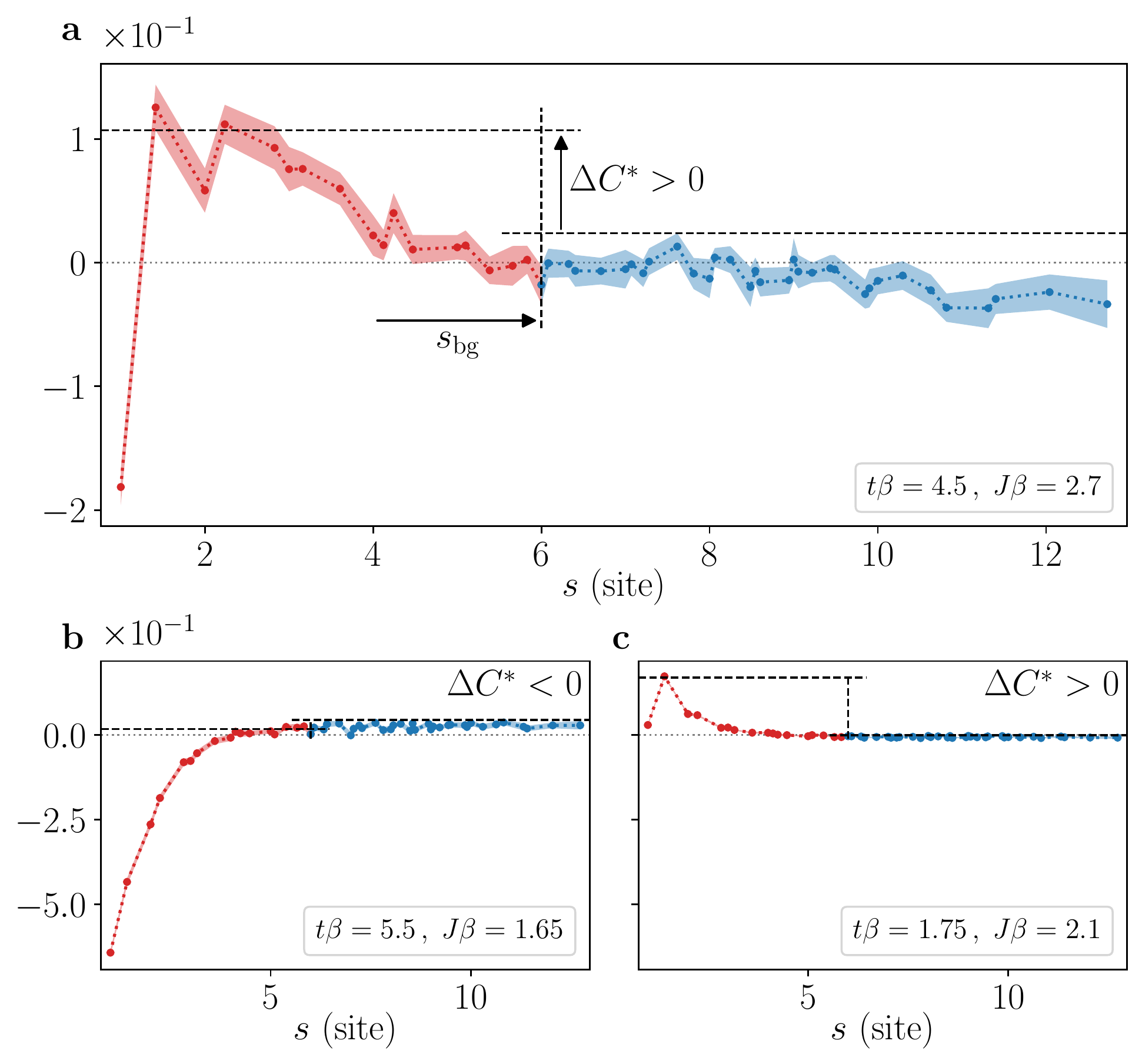}
  \caption{
    \textbf{Determining hole-hole attraction.}
    Illustrations of how hole-hole attraction is inferred using radially projected hole-hole correlations obtained using WAMC. If the difference $ \Delta C^* = \text{max} (C(s < s_\text{bg})) - \text{max} (C(s \ge s_\text{bg})) - 2 \times \text{noise} > 0 $ (\textbf{a}, \textbf{c}), the data is said to indicate attraction, otherwise (\textbf{b}) attraction is deemed to be absent.
    The partition divider $ s_\text{bg} = 6 $ is chosen such that it exceeds separation distances where a peak is observed.
  }
  \label{fig:examples}
\end{figure}

\begin{figure}[!htb]
  \includegraphics[width=\linewidth]{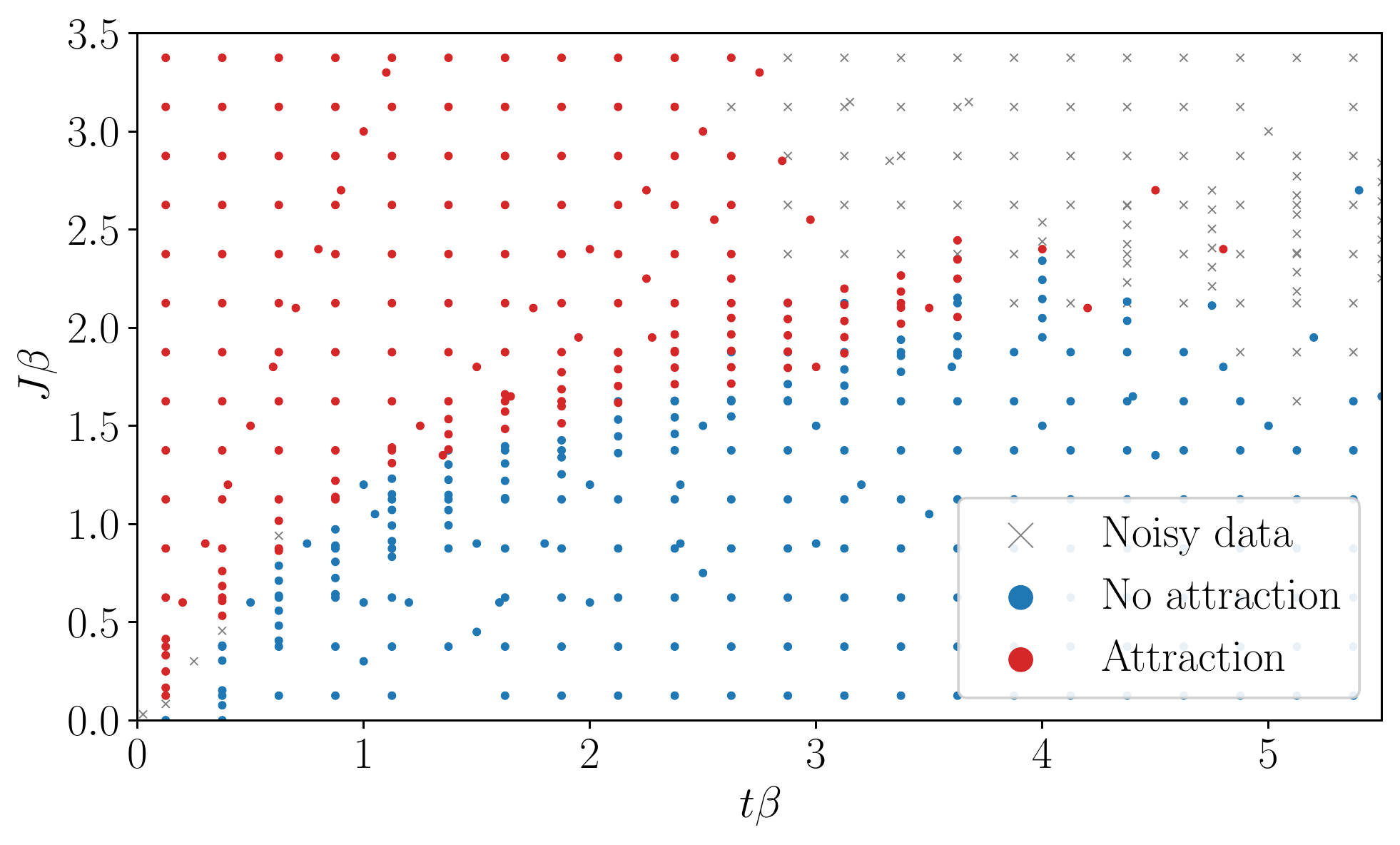}
  \caption{
    \textbf{Attractive parameter region.}
    Parameter region scanned using WAMC. Red dots indicate observed hole-hole attraction, blue dots signify no observed attraction, and a gray cross mark points deemed to have inadequate SNR. The reason for large parameter values suffering from a poor SNR can be explained by the sign problem (c.f.\ \Figref{fig:sign_map}). On the other hand, the low SNR for small parameter values is caused by a weak signal.
  }
  \label{fig:diagram}
\end{figure}

The starting point for determining hole-hole attraction is the hole-hole correlator, which in a frame of reference where one hole is always located at the origin takes the from
\begin{equation}
  C^\mathrm{h}(\mathbf{s})
  =
  (N-1)
  \big \langle
  \hat n^\mathrm{h}_{\mathbf{0}}
  \hat n^\mathrm{h}_{\mathbf{s}}
  \big \rangle - 1 \,.
  \label{Ch}
\end{equation}
The correlator is then projected onto the separation distance $ s = | \mathbf{s} | $, and the resulting radial correlation is then partitioned into two parts: the signal part in which the holes are in the close vicinity of one another $ s < s_\mathrm{bg} $, and the background part where they are further separated $ s \ge s_\mathrm{bg} $.
The condition for classifying a parameter point as attractive is that the peak value of the signal partition exceeds the background partition's peak value --- including two standard deviations of uncertainty (noise). This is illustrated in \Figref{fig:examples}, and the parameter region displaying attraction is shown in \Figref{fig:diagram}.
%
%

The value of $ s_\mathrm{bg} $ is chosen such that the attractive part of the correlation is not included in the background partition. We choose $ s_\text{bg} = 6 $, which exceeds the separation where a peak is observed.


Due to the presence of the fermionic sign problem, the noise grows exponentially with an increased value of $ t \beta $ and $ J \beta $. Data at these parameter values, therefore, suffer from a poor SNR. Data at small values of $ t \beta $ and $ J \beta $ also have a low SNR, but here the reason being a weak signal. To filter out noisy data, the SNR is defined as $ (C_\mathrm{max} - C_\mathrm{min})^2 / \langle n^2 \rangle $, i.e., the ratio of squared signal amplitude to mean square noise. A data point is deemed too noisy and ignored if $ \text{SNR} < 200 $.

To reduce noise, the outer edges of the background correlation, where $ s_x, s_y = 10 $ (we performed simulations on systems with $ 20 \times 20 $ lattice sites), are omitted. These omitted correlation values merely carry half, or even a quarter, of the statistical weight compared to other background values. The SNR of the remaining correlator is, therefore, improved.

\begin{figure}[!htb]
  \includegraphics[width=\linewidth]{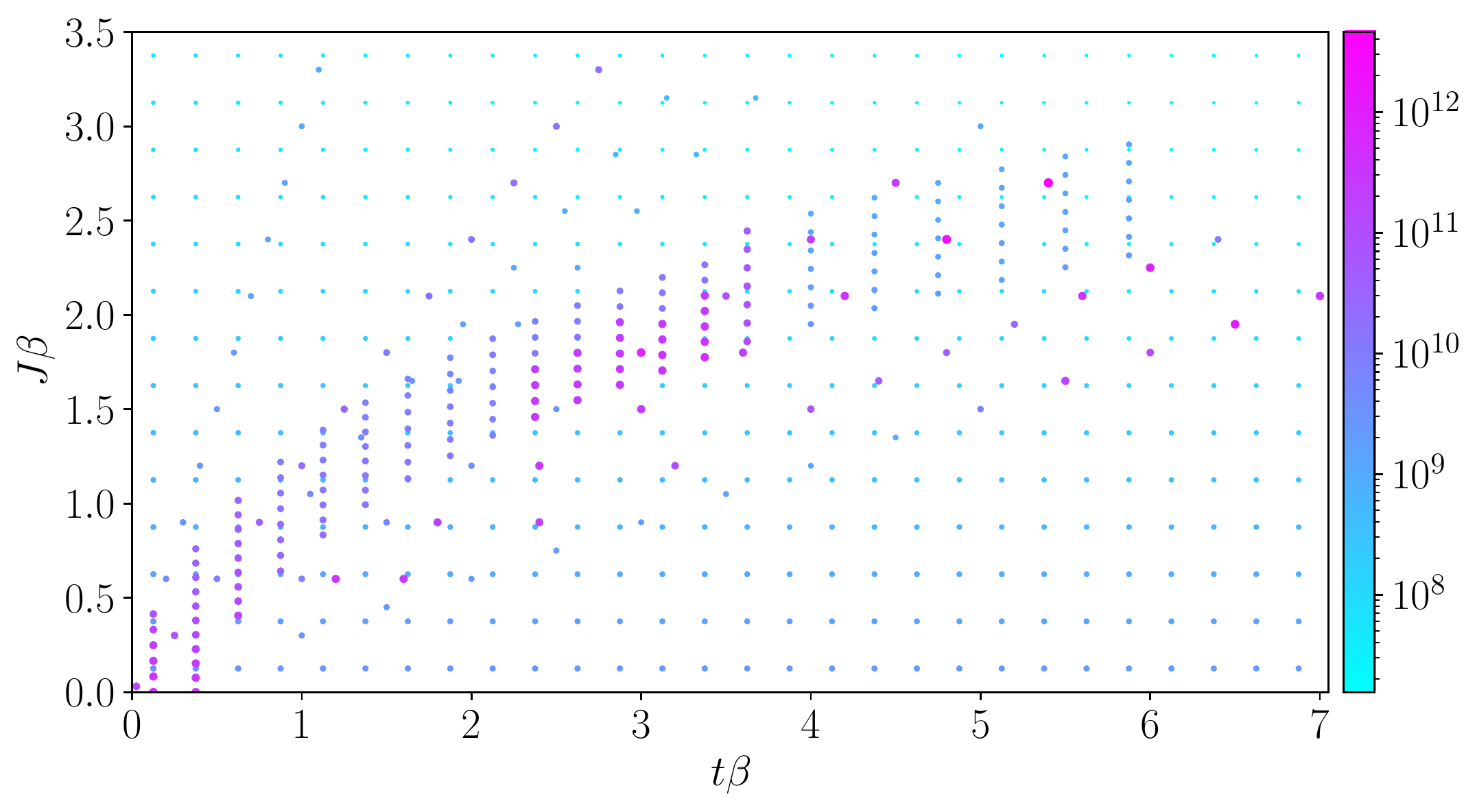}
  \caption{
    \textbf{Extent of simulation.}
    This figure illustrates the number of sampled world-line configurations in the partition function sector for different values of $ t \beta $ and $ J \beta $. The combined total simulation time amounts to several million core-hours.
  }
  \label{fig:diagram_count}
\end{figure}

To further improve SNR, we ultimately performed large-scale simulations, amounting to several million core-hours. The number of sampled world-line configurations for different values of $ t\beta $ and $ J\beta $ is shown in \Figref{fig:diagram_count}.

\subsection{Additional hole-hole correlation data}

\begin{figure}[!htb]
  \includegraphics[width=\linewidth]{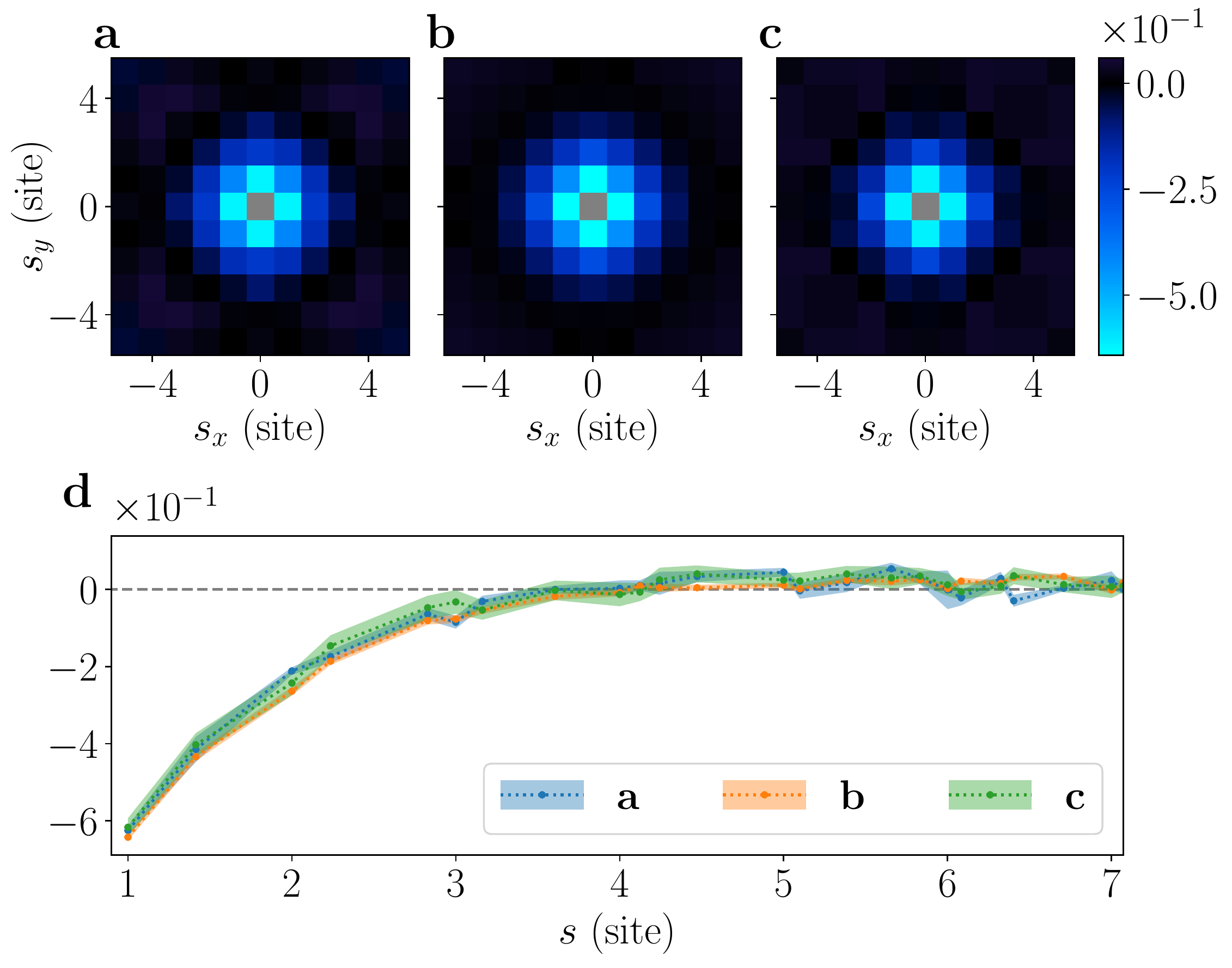}
  \caption{
    \textbf{Hole-hole correlations.}
    (\textbf{a}-\textbf{c}) show contour plots of the correlator $C^\mathrm{h}(\mathbf{s})$ for the case $ t/J = 10/3 $ and $ \beta J = 1.35,\;1.65,\;1.95$, respectively. The radial components of these are given in (\textbf{d}). No attraction is found for these parameter values.
  }
  \label{fig:hole-hole_1}
\end{figure}

\begin{figure}[!htb]
  \includegraphics[width=\linewidth]{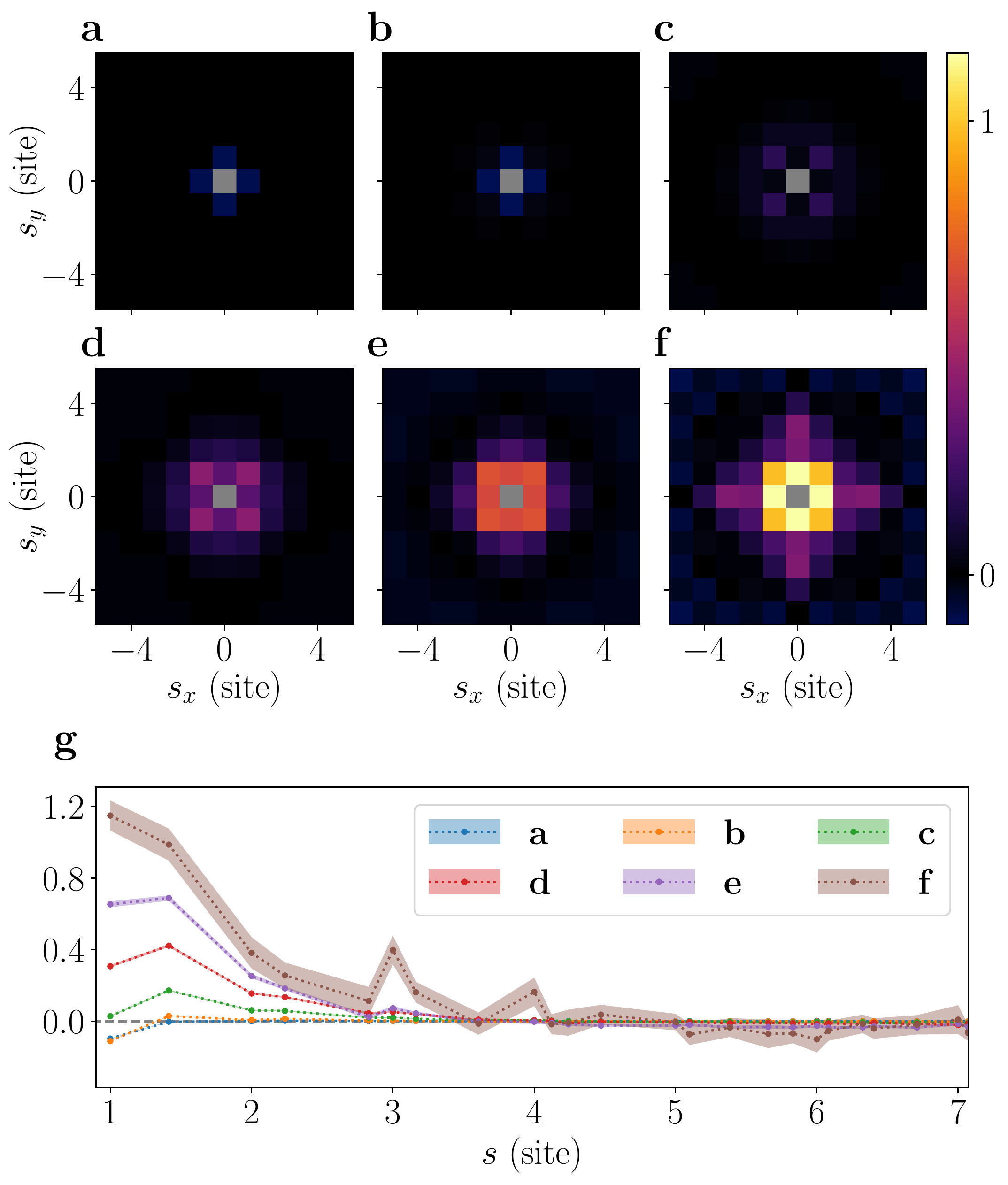}
  \caption{
    \textbf{Hole-hole correlations.}
    (\textbf{a}-\textbf{f}) show contour plots of the correlator $C^\mathrm{h}(\mathbf{s})$ for the case $ t/J = 5/6 $ and $ \beta J = 0.90,\;1.50,\;2.10,\;2.70,\;3.30,\;3.90$, respectively. The radial components of these are given in (\textbf{g}). Attraction is present for all of these parameter values.
  }
  \label{fig:hole-hole_2}
\end{figure}

In Fig.\ \ref{fig:hole-hole_1}, \ref{fig:hole-hole_2}, we present additional data of the hole-hole correlator defined in \Eqref{Ch}.

\subsection{Additional spin-spin correlation data}

\begin{figure*}[!htb]
  \includegraphics[width=\textwidth]{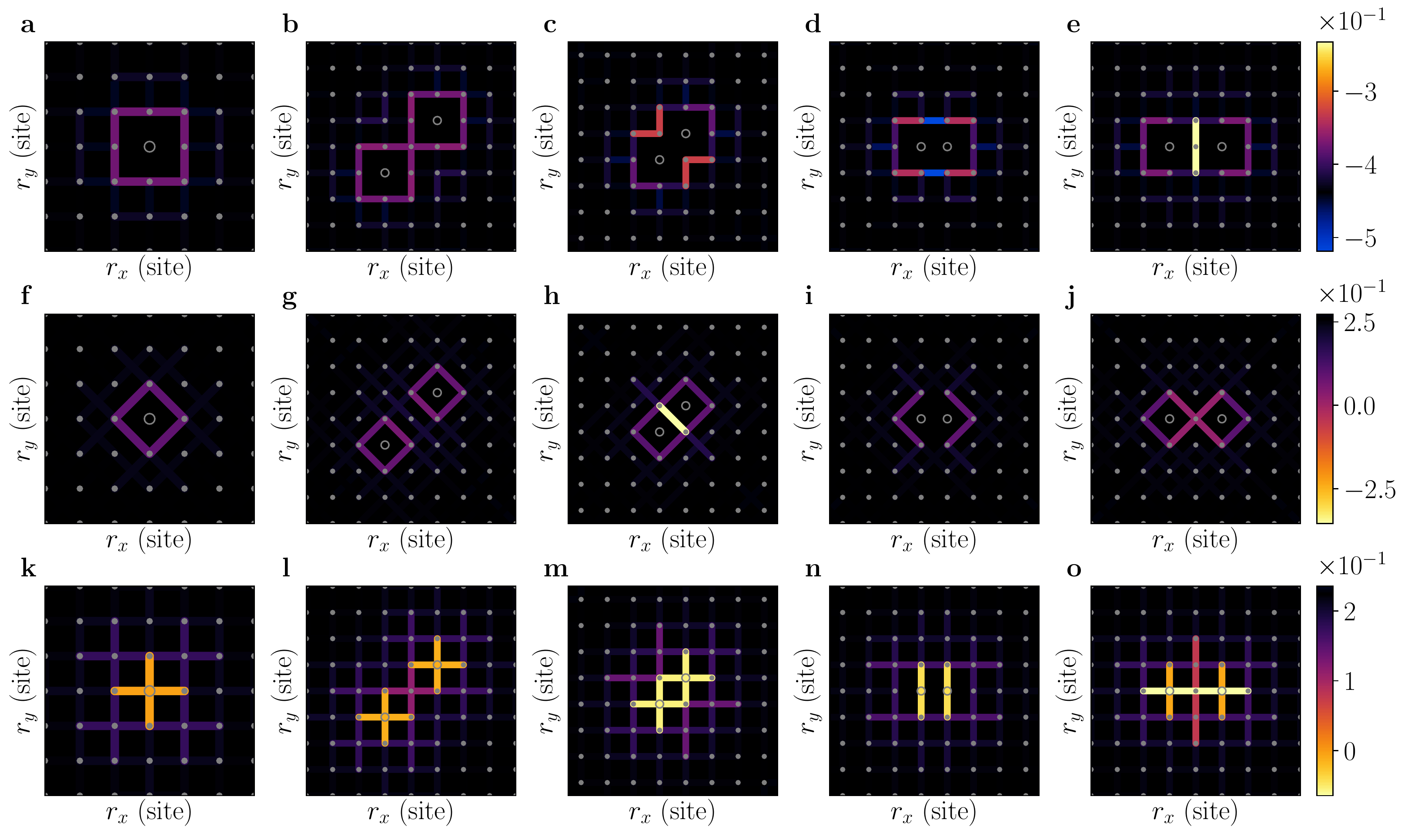}
  \caption{
    \textbf{Spin correlations $ C^\mathrm{s}_{|\mathbf d|, \mathbf s}(\mathbf r) $.}
    Left column depicts a single hole, while the remaining columns depict two dopants located in the close vicinity of one another.
    The top row (\textbf{a} - \textbf{e}) gives $ C^\mathrm{s}_{|\mathbf d|=1, \mathbf s}(\mathbf r) $, middle row (\textbf{f} - \textbf{j}) gives $ C^\mathrm{s}_{|\mathbf d|=\sqrt{2}, \mathbf s}(\mathbf r) $, and bottom row (\textbf{k} - \textbf{o}) gives $C^\mathrm{s}_{|\mathbf d|=2, \mathbf s}(\mathbf r) $.
    Model parameters are $ t \beta = 2.75 $ and $ J \beta = 3.3 $.
  }
  \label{fig:spin-spin_1}
\end{figure*}

\begin{figure*}[!htb]
  \includegraphics[width=\textwidth]{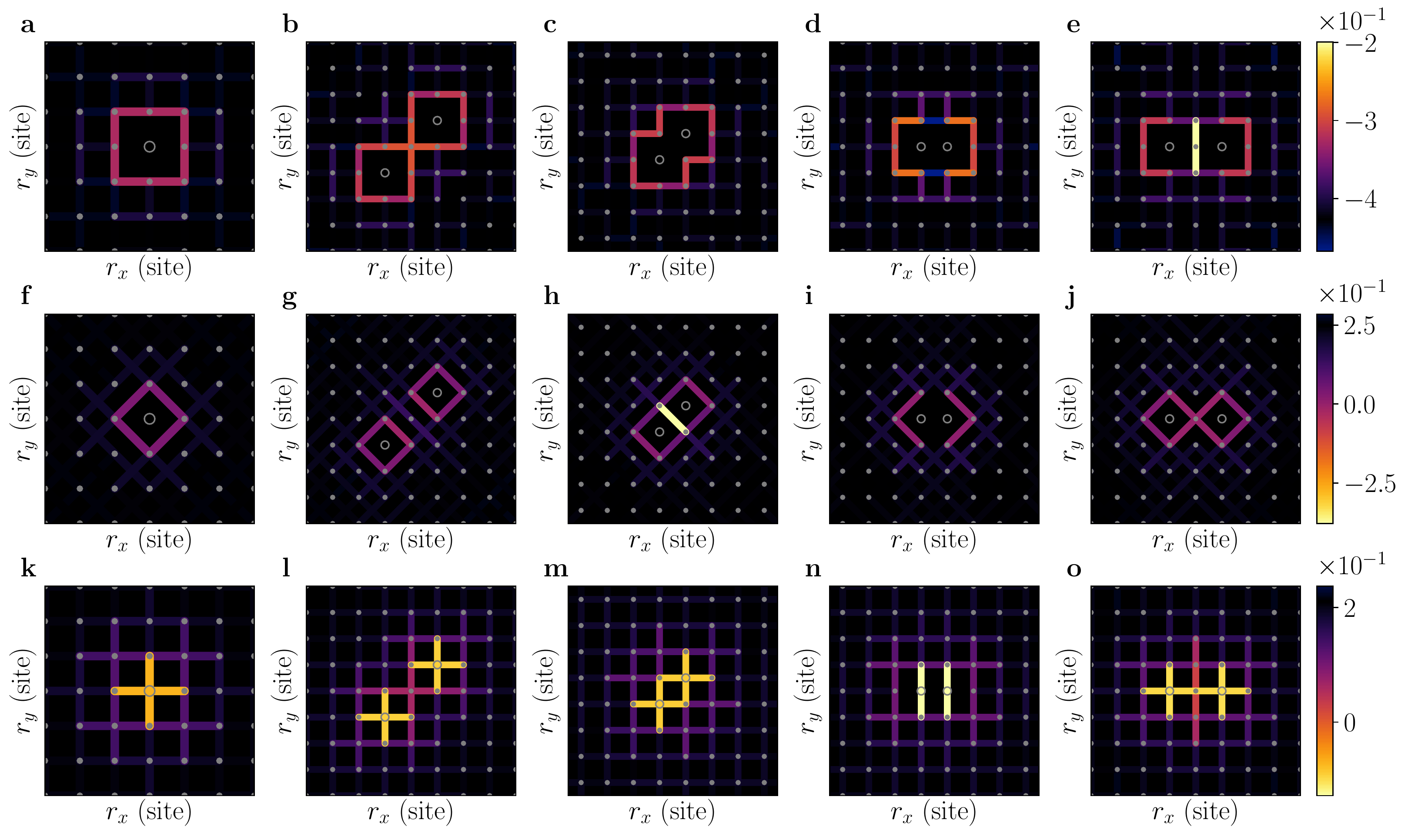}
  \caption{
    \textbf{Spin correlations $ C^\mathrm{s}_{|\mathbf d|, \mathbf s}(\mathbf r) $.}
    Left column depicts a single hole, while the remaining columns depict two dopants located in the close vicinity of one another.
    The top row (\textbf{a} - \textbf{e}) gives $ C^\mathrm{s}_{|\mathbf d|=1, \mathbf s}(\mathbf r) $, middle row (\textbf{f} - \textbf{j}) gives $ C^\mathrm{s}_{|\mathbf d|=\sqrt{2}, \mathbf s}(\mathbf r) $, and bottom row (\textbf{k} - \textbf{o}) gives $C^\mathrm{s}_{|\mathbf d|=2, \mathbf s}(\mathbf r) $.
    Model parameters are $ t \beta = 4.5 $ and $ J \beta = 2.7 $.
  }
  \label{fig:spin-spin_2}
\end{figure*}

\begin{figure*}[!htb]
  \includegraphics[width=\textwidth]{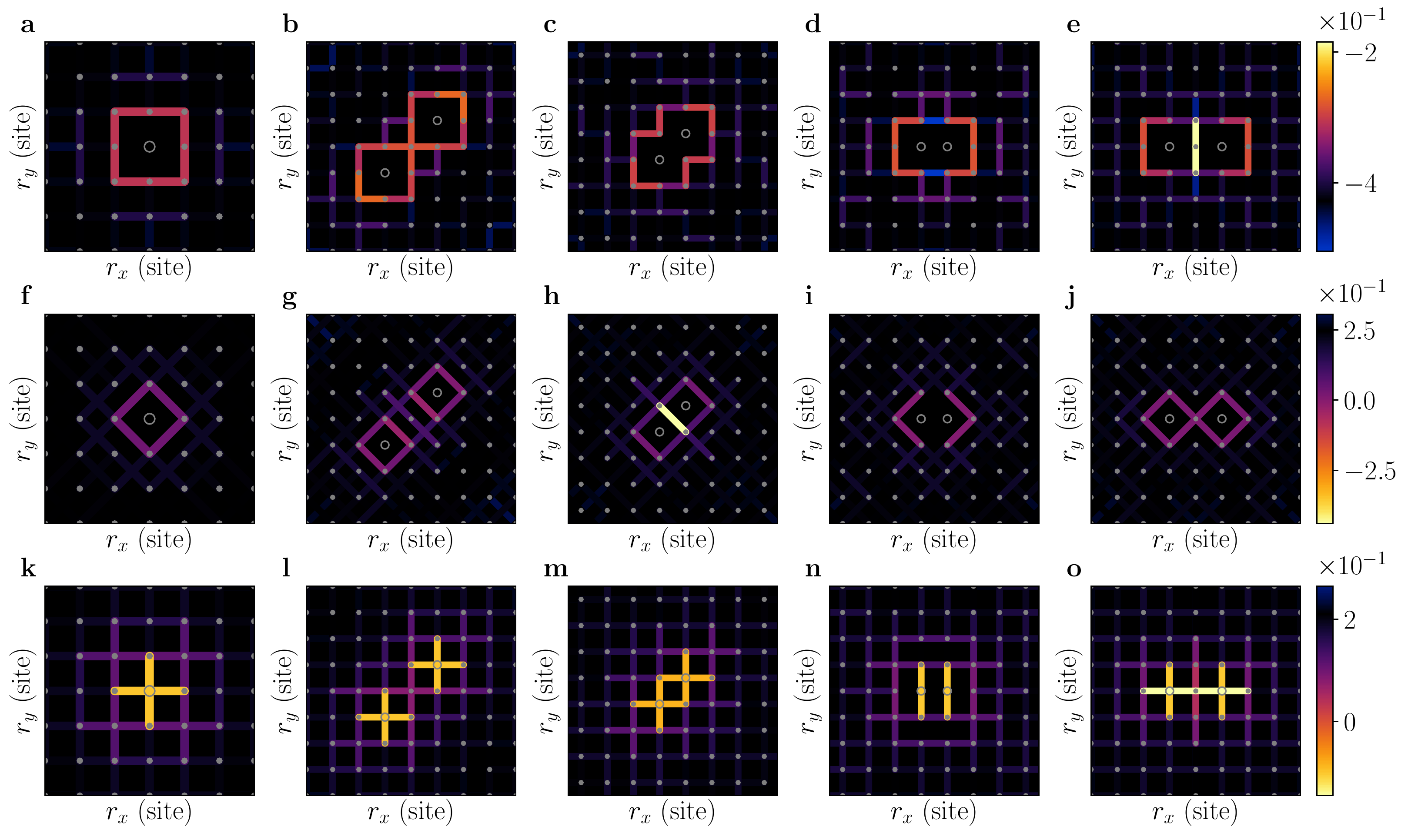}
  \caption{
    \textbf{Spin correlations $ C^\mathrm{s}_{|\mathbf d|, \mathbf s}(\mathbf r) $.}
    Left column depicts a single hole, while the remaining columns depict two dopants located in the close vicinity of one another.
    The top row (\textbf{a} - \textbf{e}) gives $ C^\mathrm{s}_{|\mathbf d|=1, \mathbf s}(\mathbf r) $, middle row (\textbf{f} - \textbf{j}) gives $ C^\mathrm{s}_{|\mathbf d|=\sqrt{2}, \mathbf s}(\mathbf r) $, and bottom row (\textbf{k} - \textbf{o}) gives $C^\mathrm{s}_{|\mathbf d|=2, \mathbf s}(\mathbf r) $.
    Model parameters are $ t \beta = 5.4 $ and $ J \beta = 2.7 $.
  }
  \label{fig:spin-spin_3}
\end{figure*}

In Fig.\ \ref{fig:spin-spin_1}-\ref{fig:spin-spin_3}, we present additional data of the spin-spin correlator, defined by
\begin{equation}
  \label{eq:spin-corr}
  C^\mathrm{s}_{\mathbf{|d|}, \mathbf{s}}(\mathbf{r})
  =
  4
  \frac{
    \sum \limits_{\mathbf{r}_0}
    \big \langle
      \hat n^\mathrm{h}_{\mathbf{r}_0}
      \hat n^\mathrm{h}_{\mathbf{r}_0 + \mathbf{s}}
      \hat S^z_{\mathbf{r}_0 + \mathbf{r} - \mathbf{d}/2}
      \hat S^z_{\mathbf{r}_0 + \mathbf{r} + \mathbf{d}/2}
    \big \rangle
  }{
    \sum \limits_{\mathbf{r}_0}
    \big \langle
      \hat n^\mathrm{h}_{\mathbf{r}_0}
      \hat n^\mathrm{h}_{\mathbf{r}_0 + \mathbf{s}}
    \big \rangle
  } \,.
\end{equation}
Further examples are given as animations \cite{youtube}.

\subsection{Exact Diagonalization}

\begin{figure}[!htb]
  \includegraphics[width=\linewidth]{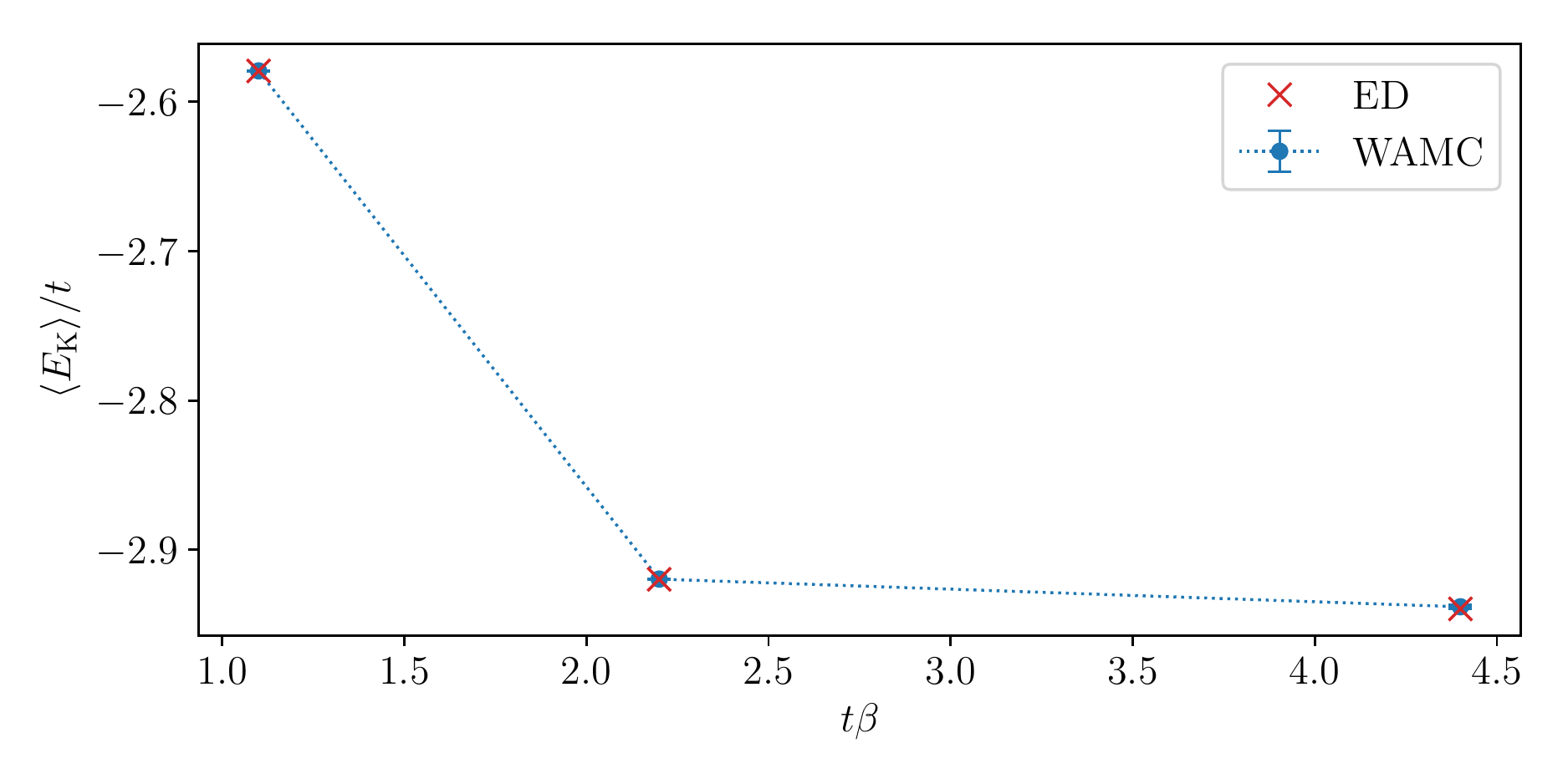}
  \caption{
    \textbf{Kinetic energy.}
    A single carrier's average kinetic energy, as a function of temperature, in a system of $ 4 \times 4 $ lattice sites with periodic boundary conditions. The ratio of hopping strength to super-exchange strength is $ t/J = 10/3 $. While error bars are given for the WAMC data, they are smaller than the marker size. The data indicate a perfect agreement between WAMC and ED.
  }
  \label{fig:E_kin}
\end{figure}

In order to verify the accuracy of WAMC, it is been benchmarked against exact diagonalization (ED) \cite{doi:10.1063/1.3518900} on a system of $ 4 \times 4 $ lattice sites with periodic boundaries, containing a single carrier. In \Figref{fig:E_kin} we present average kinetic energy results, which indicate a perfect agreement between the unbiased WAMC and exact ED method.

\bibliography{SI}